\documentclass[12pt]{article}

\usepackage[a4paper,margin=1in]{geometry}
\usepackage{setspace}
\usepackage{graphicx}
\usepackage{booktabs}
\usepackage{longtable}
\usepackage{caption}
\usepackage{subcaption}
\usepackage{amsmath}
\usepackage{natbib}
\usepackage{hyperref}
\usepackage{float}
\title{When Are Social Ties Associated with Strategic Behavior?}

\author{
Nandini Maroo\\
Cognitive Science Lab, IIIT-H\\
\texttt{nandini.maroo@research.iiit.ac.in}
\and
Kavita Vemuri\\
Cognitive Science Lab, IIIT-H\\
\texttt{kavita.vemuri@iiit.ac.in}
}

\date{}

\begin{document}
\maketitle

\begin{abstract}
Social relationships are known to shape human behavior, yet when and how social ties influence strategic cognition remains unclear. We adopt a dual-measure approach that combines observed gameplay behavior with elicitation of partner-specific beliefs at each decision point, allowing us to examine how social ties shape both decisions and predictions across interaction structures. Dyads classified as having no ties, weak ties, or strong ties played three canonical economic games: the Dictator Game, Ultimatum Game, and Centipede Game, while also making predictions about their partner's actions. Using a mixed design that held partners constant across games while varying social distance between dyads, we examined how relational proximity affected the alignment between behavior and partner-specific beliefs. Across two norm-saturated games (Dictator and Ultimatum), neither offers nor belief calibration differed reliably by social distance. In contrast, in the sequential Centipede Game, where outcomes depend on anticipating a specific partner's future actions, strong-tie dyads both cooperated longer and expected later termination than no-tie dyads, with beliefs and behavior shifting in parallel. These results indicate that social ties become strategically relevant when the interaction structure makes partner-specific accountability cognitively necessary, but not when behavior is governed primarily by shared norms or institutional constraints. The findings provide a structural account of when relational knowledge enters strategic cognition and help reconcile mixed results in prior work on social distance in economic games.
\end{abstract}


\section{Background}

Human social life is fundamentally structured by strategic interactions: situations in which individuals make decisions that depend on and influence the choices of others. From negotiating workplace agreements to coordinating household responsibilities, people routinely engage in interactions where outcomes depend not only on their own actions but also on the anticipated or actual behavior of their interaction partners. A central question in understanding these interactions concerns the role of social relationships: Do social ties between interaction partners systematically influence strategic behavior? And if so, does this influence manifest uniformly across different types of strategic situations, or does it depend on the specific structure of the interaction?

While considerable research has examined how social context affects behavior in isolated experimental paradigms, whether associations with social tie strength are structurally contingent, varying systematically across games with different strategic architectures, and whether these associations appear at the belief level in addition to the behavioral level remain open questions. The present study addresses these gaps by examining how social tie strength is associated with behavior and beliefs across three canonical economic games (the Dictator Game, Ultimatum Game, and Centipede Game) using a mixed design that allows for direct comparison across structurally different interaction types.

\subsection{Social Ties as Relational and Cognitive Structures}
Social ties vary not only in their existence but in their strength and character. Building on the seminal distinction between strong and weak ties \citep{Granovetter1973}, subsequent work has treated tie strength as a graded property indexed by differences in interaction frequency, emotional intensity, duration, and mutual obligation \citep{Marsden1984}. A frame-based approach further clarifies this conceptual space by representing a social tie as a dyadic relation endowed with a structured set of attributes and values \citep{Lizardo2024}. Within this framework, ties can be analytically distinguished into null, weak, and strong forms based on systematic differences in the values taken by strength-related attributes, while holding constant role labels, institutional settings, and categorical membership.

Within this formulation, null ties correspond to dyads that are connected through shared institutional or categorical contexts but lack a history of direct interaction \citep{Lizardo2024}. Such ties are minimally specified in the relational frame and therefore provide little basis for partner-specific inference. Weak ties occupy an intermediate position: they are associated with low values on strength-related sub-attributes such as interaction frequency, duration, and subjective closeness, typically arising from intermittent co-presence or limited shared activities. Strong ties, by contrast, are characterized by consistently high values on these same attributes, reflecting repeated intentional interaction, accumulated relational history, and affective or supportive bonds \citep{Lizardo2024, Marsden1984}. This tripartite distinction isolates variation in tie strength as an attribute of the dyad itself, rather than conflating it with differences in social roles, statuses, or institutional contexts.

These distinctions correspond to qualitative differences in the cognitive and relational frames that social-cognitive theory posits individuals deploy when representing interaction partners \citep{Brewer1988, Fiske1990}. Within this framework, the availability of relational history constrains whether others are represented at an individuated or category-based level. Strong ties satisfy the conditions under which individuated representations are theoretically expected to arise, supporting relatively stable partner-specific expectations grounded in accumulated interaction history \citep{Brewer1996}. Weak ties, by contrast, provide limited individuating information and therefore constrain representations toward thinner relational frames that rely more heavily on schematic or category-based cues \citep{Brewer1988}. In the absence of relational history, null ties constrain expectation formation to category-, role-, and institution-based representations, consistent with the default predictions of self-categorization theory \citep{Turner2012}.

Following a frame-based perspective, variation in tie strength corresponds to systematic differences in the degree of relational embeddedness and perceived similarity that structure the background social knowledge available to actors prior to interaction \citep{Lizardo2024}. Null, weak, and strong ties differ in the amount and organization of relational content they embed, thereby making different attribute–value structures salient for representing interaction partners. As a consequence, these differences in embeddedness and similarity constrain the informational basis on which beliefs and decisions are formed in interactive settings. From this view, predictable differences across interaction contexts arise not because behavior is generated anew at the point of interaction, but because expectations and coordination are anchored in pre-existing relational frames supplied by prior social relations \citep{Krackhardt1988, Uzzi1997}.

\subsection{Structural Variation in Strategic Interaction Tasks}
Strategic interactions vary along multiple structural dimensions, including the degree of outcome interdependence, the presence or absence of sequential decision-making, and the temporal extension of interaction. Three widely studied economic games capture important distinctions along these dimensions.

The \textbf{Dictator Game} represents a simple form of resource allocation, in which one player unilaterally decides how to divide an endowment between themselves and another party who has no decision-making power \citep{Forsythe1994}. Because the recipient cannot influence the outcome, the Dictator Game is commonly used to isolate distributional preferences and norm-guided behavior from strategic considerations \citep{Engel2011}. Behavior in this game is often interpreted as reflecting internalized fairness norms, altruistic motivations, or concerns related to social image, although these interpretations are not mutually exclusive \citep{Dana2006, Hoffman1994}.

The \textbf{Ultimatum Game} introduces strategic interdependence by granting the recipient veto power over the proposed allocation \citep{Gth1982}. As a result, the proposer must take into account not only their own preferences but also expectations about the responder’s likelihood of acceptance or rejection \citep{Camerer1995}. Rejection results in both parties receiving nothing, creating a tension between self-interest and the need to make offers that are perceived as acceptable \citep{Fehr1999}. This structure introduces elements of bargaining and strategic anticipation that are absent from the Dictator Game \citep{Thaler1988}.

The \textbf{Centipede Game} extends strategic interdependence across multiple sequential rounds \citep{Rosenthal1981}. Players alternate deciding whether to continue the interaction (pass) or terminate it (take), with continuation increasing the total available payoff while transferring control to the other player \citep{McKelvey1992}. Unlike the one-shot structure of the Dictator and Ultimatum Games, the Centipede Game requires players to form expectations about their partner’s future behavior across successive decision points, making cooperation contingent on beliefs about future reciprocation rather than immediate enforcement \citep{Rapoport2003}.

Taken together, these three games span a range of strategic structures: unilateral allocation (Dictator), bilateral negotiation under veto power (Ultimatum), and temporally extended strategic interaction (Centipede). This variation provides a principled basis for examining whether associations with social tie strength are uniform across interaction types or instead moderated by the strategic structure of the interaction.

\subsection{Beliefs and Mental Models in Strategic Interaction}

Strategic behavior is not determined solely by payoff structures and observable actions; it also depends on the beliefs individuals hold about their interaction partners. Psychological game theory formalizes this insight by allowing players’ utilities to depend on beliefs about others’ intentions, expectations, and actions \cite{Geanakoplos1989, Rabin1993}. Within this framework, differences in strategic behavior across social contexts may arise not only from preferences or norms, but also from differences in the content, structure, and calibration of beliefs about co-players \cite{Battigalli2007}.

A direct behavioral expression of such beliefs is the predictions individuals make about others’ actions. Social prediction integrates multiple sources of information, including category-based expectations, person-specific knowledge, and online processes such as action reading and perspective taking \cite{Frith2006, Macrae2000, Saxe2004}. Relationship history supports the accumulation of individual-specific information, allowing expectations about a particular partner to become more differentiated and less reliant on generic category-based inferences \cite{Funder1988, Paulhus1992, Friesen2011}.

Social ties are therefore expected to shape the content, differentiation, and target-specificity of beliefs about others, which may or may not translate into improved calibration depending on task structure. Interacting with familiar partners provides access to richer individual-specific knowledge, enabling predictions to rely more on learned regularities in a partner’s behavior and less on undifferentiated stereotypes or self-projection \cite{Stinson1992, Sened2024}. These predictions are further shaped by the interaction of rapid, bottom-up social cue processing and slower, top-down control processes that regulate bias and integrate contextual information \cite{Cunningham2004, Wheeler2005}.

Research in social and personality perception shows that familiarity increases the differentiation, stability, and target-specificity of expectations about others as relationship partners accumulate richer information across repeated interactions \cite{Srivastava2010, Vazire2008}. These effects reflect improved calibration to a particular individual rather than a general increase in predictive skill. Whether such refined person models translate into greater predictive calibration in strategic, incentive-laden settings, where behavior is shaped by norms, payoffs, and strategic reasoning as well as dispositions, remains an open question.

Measuring beliefs alongside behavior therefore provides a dual diagnostic for how social ties enter strategic cognition. If differences in behavior across tie categories are accompanied by differences in belief calibration or belief content, this would indicate that relational variation is reflected in the cognitive representations individuals use to model their partners. If behavior varies across tie strengths while beliefs do not, this would instead suggest that familiarity operates through other channels, such as normative expectations or social obligations \cite{Fehr2004}.

\subsection{Effects of Social Distance in Strategic Interactions: Existing Evidence and Design Limitations}

Prior research examining how social context and relational proximity influence behavior in economic games has produced nuanced and sometimes heterogeneous findings, with reported effects varying across methodological approaches and game structures. In laboratory studies that operationalize social proximity through procedural manipulations of anonymity and identifiability, such as revealing names, reducing social distance, or increasing the salience of the interaction partner, giving in the Dictator Game often increases under lower perceived distance, although these effects depend on how social presence is implemented \cite{Bohnet1999, Charness2008, Hoffman1996}. In contrast, studies that operationalize social relationships through pre-existing network ties or repeated interaction often find stronger relational effects when future interaction, reputation, or reciprocity is salient, whereas effects in strictly one-shot allocation contexts are less consistent \cite{Leider2009, Kollock1998, Murnighan1983}.

Findings in the Ultimatum Game have also been mixed. Some experimental results suggest that greater social or psychological distance can reduce responders’ sensitivity to unfair offers (e.g., fewer rejections of unfair offers when distance is higher), and that relationship valence can asymmetrically influence proposers’ behavior \cite{Vravosinos2019, Kim2013}. However, these effects are highly sensitive to framing and identity cues. Other work indicates that social distance manipulations have limited effects on Ultimatum offers under certain procedural conditions, such as revealing counterpart identity, suggesting that strategic considerations can outweigh distance effects \cite{Charness2008}. These findings are subject to several important limitations. First, many studies use between-subjects designs that may be vulnerable to individual differences in preferences or risk attitudes. Second, much of the literature has focused on individual game types rather than cross-game generalizability \cite{Engel2011}. Finally, relatively few studies have measured participants’ beliefs about partners’ likely behavior in addition to their own choices, limiting insight into the cognitive mechanisms underlying observed effects.

Finally, operationalizations of social context vary widely across studies, ranging from experimentally induced in-group/out-group categorizations \cite{Tajfel1971}, to manipulated familiarity \cite{Bohnet1999}, to naturally occurring social relationships \cite{Leider2009}. These distinct operationalizations likely engage different psychological mechanisms (such as categorical identification, informational familiarity, or affective bonding) complicating interpretation and comparison across findings \cite{Brewer1999}.

\subsection{The Present Study}

The present study addresses these limitations through a design in which dyads varying in naturally occurring tie strength complete multiple strategic games under standardized conditions, allowing between-group comparisons of tie-strength associations across structurally distinct interaction types (Dictator, Ultimatum, and Centipede Games). We examine how social tie strength is associated with both behavior and beliefs across three canonical economic games: the Dictator Game, Ultimatum Game, and Centipede Game, allowing direct comparison of tie-related differences across structurally distinct interaction types. By having all dyads experience all three games under standardized conditions, and by accounting for dyad-level variation through random effects, this design enables assessment of whether associations with tie strength are structurally contingent across different strategic architectures.

The study tests whether the association between social distance and strategic behavior depends on the strategic structure of the interaction. If associations with tie strength are uniform across interaction types, behavior and beliefs should differ consistently across all three games. If they are structurally contingent, differences may emerge selectively: for instance, in games requiring reasoning about future actions across multiple decision points (Centipede) but not in one-shot allocation tasks (Dictator, Ultimatum). By measuring beliefs concurrently with behavior, the study can assess whether any observed behavioral differences are accompanied by corresponding differences at the belief level, providing insight into whether social distance is associated with partner-specific expectations in addition to or instead of norm-based or preference-based mechanisms.
\subsubsection{Hypotheses}
Guided by a relational-frame account of social cognition and by psychological game theory, we tested two primary hypotheses concerning the role of social distance in strategic interaction.

\textbf{H1 (Beliefs $\times$ Social Distance $\times$ Structure):} Social tie strength will be associated with partner-specific beliefs when the strategic architecture of the game makes partner-specific modeling cognitively necessary.

\textbf{H2 (Behavior $\times$ Social Distance $\times$ Structure):} Social tie strength will be associated with cooperative or accommodating behavior when the strategic architecture of the game makes partner-specific accountability strategically relevant.

We do not predict uniform effects of social distance across all games; instead, we expect strategic structure to moderate whether and how social distance becomes cognitively and behaviorally relevant. 



\section{Methods}

\subsection{Social Distance Classification}
Social distance between co-players was elicited after completion of the experimental tasks using independent self-reports. Dyads were recruited from pre-existing social relationships rather than being formed randomly for the experiment, so these reports served to classify relationships that existed prior to the interaction. Participants described their relationship to their co-player in terms of prior interaction, duration of acquaintance, and social familiarity. These reports were used to assign each dyad to one of three social-distance categories: No ties (no prior personal relationship beyond shared institutional affiliation), Weak ties (prior acquaintance through shared courses or activities without sustained or close interaction), and Strong ties (close interpersonal relationships involving repeated interaction and personal familiarity).

Following standard network-theoretic distinctions \citep{Lizardo2024}, these categories
reflect differences in tie strength rather than differences in social role, demographic
similarity, or broader network position, which were held constant by design (all
participants were university students interacting in the same laboratory setting).

To ensure consistency and construct validity of the social-distance classification, self-reports were manually cross-checked across four strength-related attributes of social ties. Strong-tie dyads exhibited uniformly high values across all attributes, weak-tie dyads exhibited low values on closeness, frequency, and perceived support with more variable duration, and no-tie dyads exhibited null or near-null values across attributes. Reports were concordant within every dyad. Social distance was not experimentally manipulated and was not disclosed to participants during the task.

\subsection{Participants}
A total of 102 participants (51 dyads) were recruited from a university subject pool and participated in the study in exchange for course credit. Each participant interacted with one fixed co-player throughout the experiment, forming a dyad that remained constant across all tasks and trials.

Based on the post-task social-distance classification, dyads were classified into three social-distance categories: No ties (not at all familiar, sharing only university affiliation; 26 dyads), Weak ties (prior acquaintance through courses, clubs, or activities but no repeated intentional interaction; 10 dyads), and Strong ties (close friends with repeated social interaction and personal familiarity; 15 dyads).

\subsection{Design Overview}
The experiment employed a dyadic repeated-interaction design. Within each dyad, the
three games were played once in a fixed order to form Trial 1, and the same sequence
was then repeated to form Trial 2, with participants interacting with the same
co-player throughout. The sequence of games was counterbalanced across dyads but
held constant across trials within each dyad. Participants were not informed in advance
about the number of games, the order of games, or the number of trials. Instructions were
presented at the beginning of each round, and no information was provided about
upcoming games or future rounds.

No feedback was provided between trials, and participants did not receive feedback regarding either their co-player’s actions or the accuracy of their predictions at any stage of the experiment. This information structure ensured that each decision and prediction reflected participants’ current beliefs rather than learning from past outcomes.

Across games, the term “observer” refers to the participant who is not making the current decision and is eliciting a prediction about the co-player’s action. Within each game, participant roles alternated across rounds such that each individual experienced both decision-making and observer roles in a single trial.

\subsection{Experimental Tasks}

\subsubsection{Dictator Game}

In the Dictator Game (DG), one participant was assigned the role of allocator and received an endowment of 100 coins. The allocator decided how many coins to transfer to the other participant, who had no decision-making power in the allocation. While the allocation decision was made, the non-allocating participant (observer) predicted the number of coins the allocator would give. Roles alternated across two rounds of the DG such that each participant served once as allocator and once as observer.

\subsubsection{Ultimatum Game}

In the Ultimatum Game (UG), one participant acted as the proposer and was given an endowment of 100 coins to divide between themselves and their co-player. The co-player, acting as the responder, chose whether to accept or reject the proposed allocation. If the offer was rejected, both participants received zero coins for that round. Observers predicted both the size of the proposer’s offer and the responder’s accept-or-reject decision. Roles alternated across two rounds of the UG such that each participant served once as proposer and once as responder.

\subsubsection{Centipede Game}

The Centipede Game (CG) followed a sequential pass-or-take structure. The game began with a pot of 100 coins, and the initial decision-maker was randomly chosen. On each turn, the acting player decided whether to pass the pot to their co-player or to take it. Passing doubled the total pot and transferred control to the other player. Taking immediately terminated the game and divided the current pot according to a fixed 65:35 split, with the acting player receiving the larger share. At each turn, participants were shown the payoff shares they would receive if the pot were taken at that point.

Each trial contained a single CG interaction. The maximum number of rounds was five, but this limit was not disclosed to participants, and no information about the end of the game was ever provided. Observers predicted at each decision point whether their co-player would pass or take.

\subsection{Procedure and Setup}

The study was approved by the Institutional Review Board. All participants provided informed consent prior to participation. Participants were informed about the nature of the experiment, their right to withdraw, and how their data would be used.

The experiment was conducted in a laboratory setting. Members of each dyad were seated in the same room but separated by an opaque partition that prevented visual and auditory communication during the task. Participants briefly saw each other before the experiment began, after which all interactions were mediated through a computer interface. All decisions and predictions were recorded using custom-built experimental software. Participants were shown only their own coin balance at any given point and were not informed of their co-player’s cumulative earnings.

\subsection{Prediction Elicitation}
Predictions were elicited concurrently with decision-making in all games. In the DG and UG, participants provided interval-valued estimates of the number of coins their co-player would offer. In the UG and CG, participants predicted at each decision node whether their co-player would accept or reject/pass or take.

Following psychological game theory, beliefs about a co-player’s actions were treated as psychologically relevant state variables. Eliciting predictions at each decision point allowed these belief states to be measured directly rather than inferred from behavior alone. Participants were shown their own cumulative coin totals throughout the experiment, ensuring that predictions and decisions were made in a payoff-salient context even in the absence of monetary incentives. No feedback was given about prediction accuracy.

\subsection{Measures}

\subsubsection{Behavioral Measures}

Behavioral measures included offer size in the Dictator and Ultimatum Games, acceptance or rejection decisions in the Ultimatum Game, and the number of rounds completed before termination in the Centipede Game.

\subsubsection{Belief Measures}

Belief measures consisted of predicted offer intervals in the Dictator and Ultimatum Games, predicted acceptance or rejection decisions in the Ultimatum Game, and predicted pass-or-take decisions in the Centipede Game.

\subsubsection{Prediction Error}
Prediction error was defined only for offer predictions in the Dictator and Ultimatum Games. For each interval-valued prediction, the midpoint of the predicted range was computed. Prediction error was then calculated as the absolute difference between the realized offer and this midpoint.

Theoretically, this quantity captures the mismatch between a participant’s subjective belief about their co-player’s behavior and the objective social outcome that was realized.

\subsection{Data Structure and Statistical Analysis}
Observations were nested within dyads and participants, reflecting the repeated and reciprocal nature of the interactions across games and trials. Because multiple observations were collected from the same individuals and the same dyads, all analyses used mixed-effects models to account for this non-independence. Models included random intercepts for participants and dyads to account for non-independence in the nested data structure, with social distance (No ties, Weak ties, Strong ties) specified as a fixed effect. Separate models were estimated for each game type. This analytical approach allows between-group comparison of social distance effects within each game while controlling for baseline dyad-level and participant-level variation through random effects. Residuals showed mild deviations from homoscedasticity for prediction-error models, but offer and cooperation models satisfied standard assumptions, and mixed-effects estimates are robust to such deviations. All analyses were conducted in Python using the \texttt{statsmodels} package.

\section{Results}

All analyses were conducted using mixed-effects models with random intercepts for participants and dyads (or rooms for the Centipede Game). Separate models were estimated for each game to assess between-group differences in social distance effects. Models were estimated using restricted maximum likelihood. Social distance (No ties, Weak ties, Strong ties) was included as a fixed effect, with No ties as the reference category.

Preliminary models including trial number and its interaction with social distance revealed no reliable effects for any outcome (all $p > .10$). Trial was therefore treated as a repeated episode within dyads in all subsequent analyses.

\begin{figure}[H]
\centering
\includegraphics[width=4.5in]{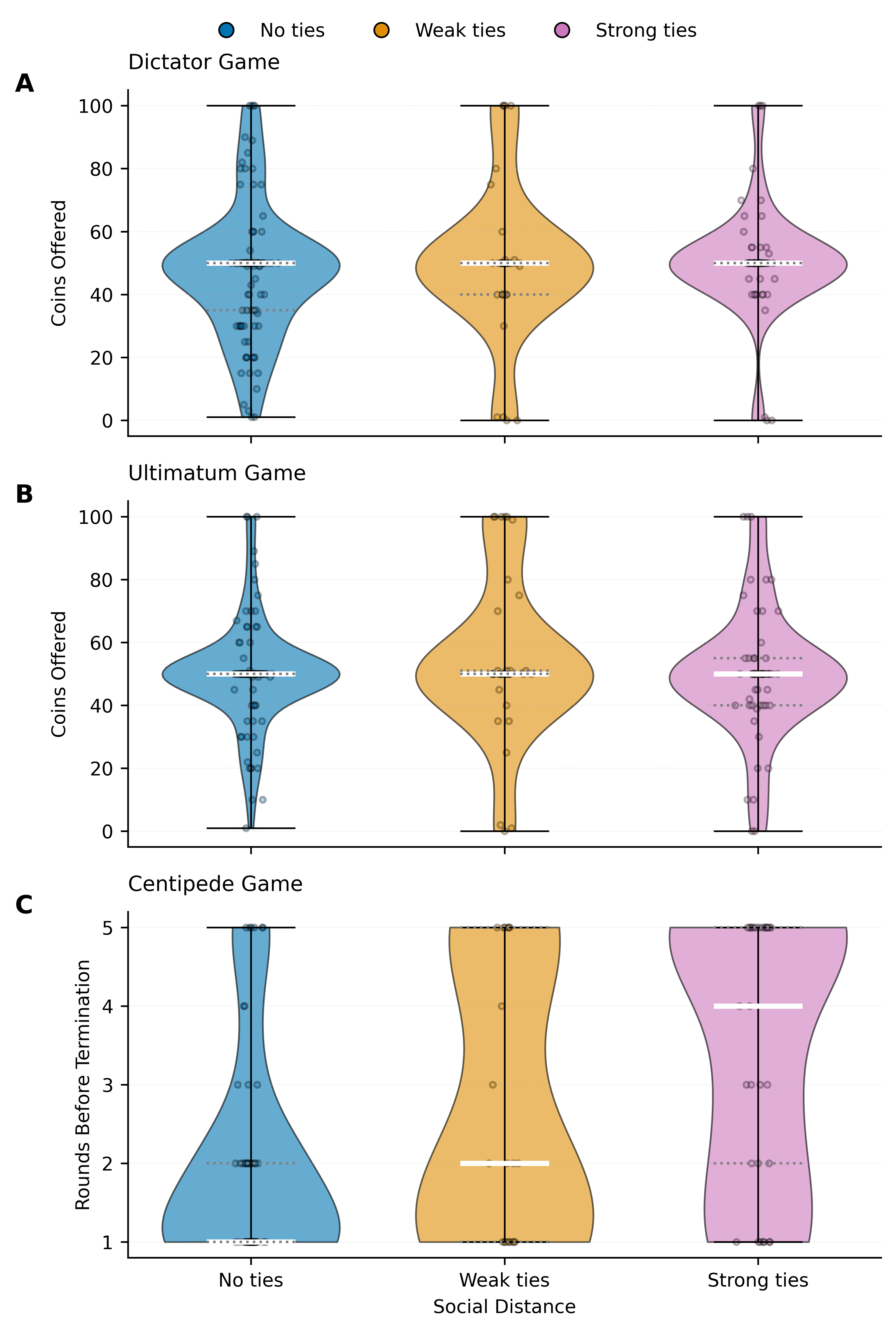}
\caption{
Behavioral outcomes by social distance.
(A) Dictator Game offer distributions.
(B) Ultimatum Game offer distributions.
(C) Number of rounds completed in the Centipede Game. (The maximum number of rounds was capped at five, producing a ceiling for some strong-tie dyads, which likely attenuates rather than inflates observed differences.)
Points represent individual observations; violins depict distributional density.
}
\label{fig:main_results}
\end{figure}
\subsection{Dictator and Ultimatum Games: Offers}

Figure~\ref{fig:main_results}A–B show offer distributions.

In the Dictator Game, baseline offers in no-tie dyads averaged 47.10 coins. Strong-tie dyads offered 3.47 coins more (95\% CI $[-4.53,\,11.47]$, $p=.395$) and weak-tie dyads 2.60 coins more (95\% CI $[-6.57,\,11.78]$, $p=.578$). Thus even moderately large relational effects on DG offers are ruled out.

In the Ultimatum Game, no-tie dyads offered 49.73 coins on average. Strong-tie dyads differed by $-0.05$ coins (95\% CI $[-7.33,\,7.24]$, $p=.990$) and weak-tie dyads by $+4.29$ coins (95\% CI $[-4.07,\,12.65]$, $p=.314$), again indicating tight null effects.

Responder acceptance rates were near ceiling across all social distances ($>90\%$). Logistic mixed-effects models showed no evidence that acceptance increased for socially closer dyads; coefficients were slightly negative for both strong and weak ties.

\subsection{Belief calibration in the Dictator and Ultimatum Games}

In the Dictator Game, prediction error for no-tie dyads averaged 19.03 coins. Strong ties differed by $-4.88$ coins (95\% CI $[-11.45,\,1.69]$, $p=.146$) and weak ties by $+3.50$ coins (95\% CI $[-4.05,\,11.04]$, $p=.364$).

In the Ultimatum Game, baseline error was 15.26 coins. Strong ties differed by $+3.56$ coins (95\% CI $[-2.58,\,9.69]$, $p=.256$) and weak ties by $+2.09$ coins (95\% CI $[-4.95,\,9.13]$, $p=.561$).

Thus, across both norm-saturated games, belief calibration shows no reliable or practically meaningful association with social distance.

\subsection{Centipede Game: Cooperation}

Figure~\ref{fig:main_results}C shows number of rounds completed in the game.

No-tie dyads terminated after 1.96 rounds on average. Strong-tie dyads continued 1.47 rounds longer (95\% CI $[0.57,\,2.38]$, $p=.001$). Because several strong-tie dyads reached the five-round cap, this estimate is conservative. Weak-tie dyads showed a smaller, statistically uncertain difference of $0.69$ rounds (95\% CI $[-0.35,\,1.73]$, $p=.194$).

\subsection{Centipede Game: Observer beliefs}

Observers in no-tie dyads expected their partner to take at round 1.93. In strong-tie dyads, this expectation occurred 1.14 rounds later (95\% CI $[0.46,\,1.83]$, $p=.001$). Weak-tie dyads differed by $0.55$ rounds (95\% CI $[-0.24,\,1.34]$, $p=.171$).

\subsection{Belief-behavior coupling in the Centipede Game}

Across dyads, observer expectations significantly predicted actual termination (mixed-effects model: $\beta = 0.124$, 95\% CI $[0.016,\,0.232]$, $p = .024$). Strong-tie dyads also remained significantly more cooperative after controlling for observer expectations ($\beta = 1.33$, $p = .001$).

Correlations between observer expectations and termination round were high in all groups (No ties: $r = .85$; Weak ties: $r = .82$; Strong ties: $r = .65$). Strong-tie dyads exhibited later expected termination points and later realized termination points than no-tie dyads.

\paragraph{Multiple-comparison correction.}
Because each game instantiates a distinct theoretical mechanism (norm-governed allocation, norm-constrained bargaining, and accountability-based sequential interaction), Holm correction was applied within each game family rather than across all outcomes. Applying a single familywise correction across all games would conflate theoretically independent hypotheses and bias against detecting structure-specific effects. Under this theory-consistent correction, both strong-tie effects in the Centipede Game (behavior and beliefs; Holm-adjusted $p=.0043$ and $p=.0042$) remain significant. No Dictator or Ultimatum Game effects approach significance after correction.

\section{Discussion}

This study examined whether naturally occurring social ties are associated with strategic
behavior and partner-specific beliefs in a way that depends on the structure of the
interaction. Across three canonical games, the results reveal a clear pattern: variation
in social tie strength is not associated with uniform differences across tasks, but becomes
consequential when the game makes forward-looking accountability to a particular partner
cognitively relevant. These results provide support for the hypothesized structural contingency for both belief- and behavior-based predictions: social distance influenced beliefs and gameplay only when the strategic architecture of the game made partner-specific modeling cognitively necessary, and not when norm-based mechanisms dominated.

In the Dictator Game (DG) and Ultimatum Game (UG), neither offers nor
belief calibration differed detectably across no-tie, weak-tie, and strong-tie dyads. In
contrast, in the Centipede Game (CG), strong-tie dyads both cooperated for longer and
expected their partners to continue for longer than no-tie dyads. Together, these findings
support a structural-contingency account: relational proximity matters when the
interaction requires accountability to a specific partner, but not when outcomes are
primarily governed by shared norms or institutionalized bargaining conventions.

Across the three games, the relevant cognitive representation changes. In the Dictator and Ultimatum Games, decisions are governed by norm-based rules (“what is a fair or acceptable offer?”), which do not require modeling a specific partner. In the Centipede Game, by contrast, decisions require forward-looking representation of a particular individual’s future actions (“what will this person do next?”). Social ties matter only in the latter regime because only there is a partner model computationally necessary.

The absence of relational differences in the DG and UG is consistent with their
interpretation as norm-saturated games. In the DG, allocators face no strategic
interdependence, so offers largely reflect fairness and generosity norms rather than
expectations about a particular partner \cite{Fehr1999, Camerer2003}. Prior reports of higher generosity toward socially closer others in DGs \cite{Charness2008, Leider2009} often rely on between-subject designs or explicit social manipulations that make social distance salient at the point of decision. The present design, in which all dyads experienced all games under standardized conditions and dyad-level variation was accounted for through random effects, shows that normative constraints remain dominant even in the presence of strong social ties.

The UG similarly embeds strong fairness norms through the responder’s veto power.
Proposers must make offers that clear an implicit acceptability threshold, and responders’
decisions are known to be guided by inequity aversion and shared norms rather than
idiosyncratic partner models \cite{Fehr1999, Camerer2003}. Consistent with this, neither
offers nor belief calibration varied with social distance. The narrow confidence intervals
suggest that any relational effects, if present, are likely to be small, reinforcing the view
that bargaining under shared norms constrains the extent to which partner-specific
knowledge can influence behavior.

The CG, by contrast, is organized around forward-looking accountability: each player’s
optimal choice depends on what a particular partner is expected to do next. In this
setting, strong-tie dyads terminated later and observers in those dyads expected later
termination points. These converging behavioral and belief-level differences indicate that stronger relational frames made more partner-specific expectations cognitively available and task-relevant in a setting where outcomes hinge on anticipating a particular individual’s future actions, rather than on generic norms or role-based scripts. This
pattern reflects shifts in the content and calibration of partner models, not greater
strategic optimality or norm-free predictive accuracy. This aligns with frame-based
theories of social cognition, which hold that strong ties support more individuated
relational representations than weak or null ties \cite{Brewer1988, Fiske1990, Lizardo2024}.

These findings integrate insights from psychological game theory and social cognition.
Psychological games emphasize that utilities depend on beliefs about others’ intentions
and actions \cite{Rabin1993, Battigalli2007}. The present results specify when such
belief-dependent reasoning becomes socially structured, namely when interaction
requires accountability to a particular partner. From a relational-frame perspective
\cite{Lizardo2024}, strong ties make partner-specific attribute–value structures available,
but whether these structures are used depends on whether the task demands individuated
representation. When accountability is low, even strong ties leave actors relying primarily
on generic norms and institutional scripts \cite{Granovetter1973, Brewer1988}.

This interpretation is further supported by the belief–behavior coupling in the CG.
Observers’ expectations about when their partner would take significantly predicted
actual termination, and strong-tie dyads remained more cooperative even after controlling
for these expectations. Thus, relational proximity shaped both the content of
partner-specific expectations and the way those expectations were translated into
action. In norm-governed games such dual pathways are suppressed, but in
accountability-based interaction they become jointly operative.

Taken together, the pattern across games highlights a sharp contrast between
norm-governed and accountability-based interaction. Norms compress behavioral variance
and render relational knowledge largely irrelevant, as seen in the DG and UG.
Accountability-based structures, as in the CG, expand the relevance of relational frames,
allowing social ties to shape both what is expected and what is done. This offers a unified
explanation for why prior work has sometimes found strong social-distance effects (e.g.,
in repeated or sequential games) and sometimes not (e.g., in one-shot
allocations): it is the strategic architecture, not relational proximity alone, that
determines when relational cognition becomes consequential.

\section{Conclusion}

This study shows that the association between social tie strength and strategic behavior is structurally contingent rather than uniform. Social proximity was associated with both behavior and beliefs in the Centipede Game, where sequential decision-making requires anticipation of a specific partner’s future actions, but not in the Dictator or Ultimatum Games, where behavior is largely constrained by norms and institutionalized bargaining rules. Relational cognition becomes consequential when the interaction structure makes partner-specific expectations strategically relevant.

A central implication of these findings is that social relationships influence strategic behavior through their effects on belief formation and prediction, rather than through preferences or norms alone. Social ties mattered precisely when participants were required to generate and rely on expectations about a particular partner’s future actions. By jointly measuring beliefs and behavior, the present study shows that relational proximity shapes not only what people do but also what they expect others to do, and that this coupling becomes consequential only under conditions of sequential accountability.

Together, these findings specify boundary conditions for when social distance enters strategic cognition. They suggest that relational representations are most likely to shape behavior in settings that require ongoing accountability to a particular partner, rather than in one-shot or norm-governed exchanges. This framework also provides a way to interpret mixed results in the literature on social distance in economic games, by locating relational influences in the structure of the task rather than in tie strength alone.

Important qualifications temper these interpretations. Social ties were naturally occurring rather than experimentally manipulated and were elicited after the interaction, so unobserved correlates of friendship or retrospective consistency effects may contribute to the observed patterns. However, any misclassification of social distance would be expected to attenuate rather than inflate relational effects, making the strong-tie differences observed in the Centipede Game conservative estimates. The laboratory setting, university sample, and limited number of trials constrain generalizability. The absence of feedback precludes analysis of learning or adaptation over time, and belief elicitation itself may have influenced behavior by increasing the salience of expectations during decision making. In the Centipede Game, termination is a single event per trial, so the effective sample size is smaller than in the Dictator and Ultimatum Games, limiting sensitivity to detect weaker relational effects such as differences between weak and no ties. In addition, the five-round cap produces a ceiling effect for some strong-tie dyads, which likely attenuates rather than exaggerates observed differences.

More broadly, the study contributes to theoretical integration between network science, psychological game theory, and social cognition by showing that social structure influences strategic interaction in a conditional way, depending on whether the task architecture privileges partner-specific accountability over generic norm application.

Future work could increase the number of interaction rounds, introduce feedback to examine the dynamics of accountability, manipulate relational context experimentally, and vary normative frames to test when institutionalized expectations suppress relational influences. Such extensions would allow a sharper characterization of how and when social relationships shape strategic reasoning.
\section*{Declarations}
The authors declare no conflicts of interest.

\section*{Data Availability}
De-identified data and analysis scripts will be made available upon request.



\bibliographystyle{unsrt}
\bibliography{cite}

\end{document}